\documentclass[12pt, onecolumn]{IEEEtran}

\usepackage{amsmath}
\usepackage{amsfonts}
\usepackage{amssymb}
\usepackage{epsfig}
\usepackage{graphicx}

\newtheorem{theorem}{Theorem}
\newtheorem{lemma}{Lemma}

\begin{document}

\title{Channel Code Design with Causal Side \\ \vspace{-8pt}
Information at the Encoder}

\author{\authorblockN{Hamid Farmanbar, Shahab Oveis Gharan, and Amir
K. Khandani\\}
\authorblockA{Coding and Signal Transmission Laboratory\\
\vspace{-7pt}
Department of Electrical and Computer Engineering\\
\vspace{-7pt}
University of Waterloo\\
\vspace{-7pt}
Waterloo, Ontario, N2L 3G1\\
\vspace{-7pt}
Email: \{hamid,shahab,khandani\}@cst.uwaterloo.ca}
\thanks{This work was presented in part at the IEEE Canadian
Workshop on Information Theory, Edmonton, Alberta, Canada, June 6-8,
2007.}}

\maketitle

\begin{abstract}
The problem of channel code design for the $M$-ary input AWGN
channel with additive $Q$-ary interference where the sequence of
i.i.d. interference symbols is known causally at the encoder is
considered. The code design criterion at high SNR is derived by
defining a new distance measure between the input symbols of the
Shannon's \emph{associated} channel. For the case of binary-input
channel, i.e., $M=2$, it is shown that it is sufficient to use only
two (out of $2^Q$) input symbols of the \emph{associated} channel in
the encoding as far as the distance spectrum of code is concerned.
This reduces the problem of channel code design for the binary-input
AWGN channel with known interference at the encoder to design of
binary codes for the binary symmetric channel where the Hamming
distance among codewords is the major factor in the performance of
the code.
\end{abstract}


\section{Introduction}
Information transmission over channels with known interference at
the transmitter has recently found applications in various
communication problems such as digital watermarking \cite{Chen01}
and broadcast schemes \cite{Caire03}. A remarkable result on such
channels was obtained by Costa, who showed that the capacity of the
additive white Gaussian noise (AWGN) channel with additive Gaussian
i.i.d. interference where the sequence of interference symbols is
known non-causally at the transmitter is the same as the capacity of
the AWGN channel \cite{CosIT83}. Therefore, the Gaussian
interference does not incur any loss in the capacity. This result
was extended to arbitrary (random or deterministic) interference in
\cite{Erez05} by using a precoding scheme based on multi-dimensional
lattice quantization. Following Costa's ``Writing on Dirty Paper"
famous title \cite{CosIT83}, coding for the channel with
non-causally known interference at the transmitter is referred to as
``dirty paper coding" (DPC). By analogy, coding for the channel with
causally-known interference at the transmitter is sometimes referred
to as ``dirty tape coding" (DTC). The result obtained by Costa does
not hold for the case that the sequence of interference symbols is
known causally at the transmitter.

Recently, dirty paper coding has emerged as a building block in
multiuser communication. In particular, there has been considerable
research studying the application of dirty paper coding to broadcast
over multiple-input multiple-output (MIMO) channels. In such
systems, for a given user, the signals sent to other users are
considered as interference. Since all signals are known to the
transmitter, successive ``dirty paper" cancelation can be used in
transmission after some linear preprocessing \cite{Caire03}. It was
shown that DPC in fact achieves the sum capacity of the MIMO
broadcast channel \cite{Yu04, Gold03, Tse03}. Most recently, it has
been shown that the same is true for the entire capacity region of
the MIMO broadcast channel \cite{Wein06}.

These developments motivate finding realizable dirty paper coding
techniques. Building upon \cite{Erez05}, Erez and ten Brink
\cite{Erez06} proposed a practical code design based on vector
quantization via trellis shaping and using powerful channel codes.
Due to the complexity of implementation, their scheme uses the
knowledge of interference up to six future symbols rather than the
whole interference sequence. Bennatan \textit{et al.} \cite{Ben06}
gave another design based on superposition coding and successive
cancelation decoding. Their design uses a trellis coded quantizer
with memory length nine and a low density parity check (LDPC) code
as channel code. Wei Yu \textit{et al.} \cite{Yu05} gave a design
based on convolutional shaping and channel codes.

The schemes that use the interference sequence up to the current
symbol can be used as low-complexity solutions for the dirty paper
problem. For example, in \cite{Chen01}, scalar lattice quantization
is proposed for data-hiding even though in that context, the host
signal in clearly known non-causally.

In this paper, we consider the problem of channel code design for
the $M$-ary input AWGN channel with additive causally-known discrete
interference. The discrete interference model is more appropriate
for many practical applications. For example, in the MIMO broadcast
channel where the transmitter uses a finite constellation, the
interference caused by other users is discrete rather than
continuous.

Our design does not rely on the suboptimal (in terms of capacity)
precoding scheme based on scalar lattice quantization for the dirty
tape channel \cite{Erez05}, \cite{Caire02}. Instead, we consider a
\emph{new} approach based on code design for the Shannon's
\emph{associated} channel over all possible input symbols. Another
distinction between our work and the related research in the field
is that we consider a finite channel input alphabet rather than a
continuous one.

This paper is organized as follows. In the next section, we
summarize Shannon's work on channels with causal side information at
the transmitter. In section \ref{chanmodel}, we introduce the
channel model. In section \ref{code}, we derive the code design
criterion for the AWGN channel with causally-known discrete
interference at the encoder. In section \ref{binary}, we consider
channels with binary input for which we show that the design
criterion derived in section \ref{code} reduces to maximizing the
Hamming distance. In section \ref{Mary}, we consider a special case
for which the result for the binary channel also holds for the
$M$-ary channel. In section \ref{general}, we consider a more
general channel model for which the main results of this work hold.
We conclude this paper in section \ref{conclude}.

\section{Channels with Side Information at the Transmitter}
\label{SI}
Channels with known interference at the transmitter are special case
of channels with side information at the transmitter which were
considered by Shannon \cite{Shan58} in the causal knowledge setting
and by Gel'fand and Pinsker \cite{Gel80} in the non-causal knowledge
setting.

\begin{figure}[t]
\centering
\includegraphics[scale=0.9]{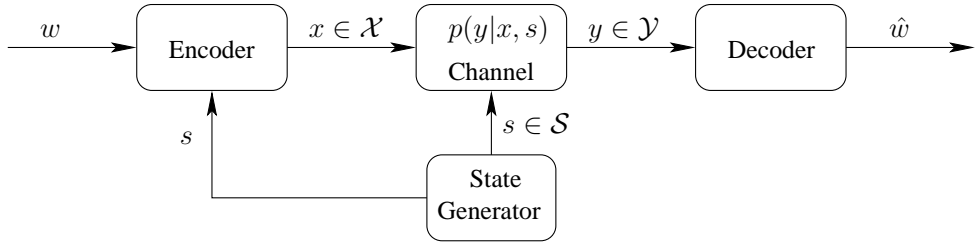}
\caption{SD-DMC with state information at the encoder.}
\label{J2-fig1}
\end{figure}

Shannon considered a discrete memoryless channel (DMC) whose
transition matrix depends on the channel state. A state-dependent
discrete memoryless channel (SD-DMC) is defined by a finite input
alphabet $\mathcal{X}$, a finite output alphabet $\mathcal{Y}$, and
transition probabilities $p(y|x,s)$, where the state $s$ takes on
values in a finite alphabet $\mathcal{S}$. The block diagram of a
state-dependent channel with state information at the encoder is
shown in fig. \ref{J2-fig1}.

In the causal knowledge setting, the encoder maps a message $w$ into
$\mathcal{X}^n$ as
\begin{equation}
x_i = f_i\left(w,s_1,\ldots,s_i\right), \quad 1 \leq i \leq n.
\end{equation}

Shannon showed that it is sufficient to consider the coding schemes
that use only the current state symbol in the encoding process to
achieve the capacity of an SD-DMC with i.i.d. state sequence known
causally at the encoder \cite{Shan58}.

\begin{figure}[t]
\centering
\includegraphics[scale=1]{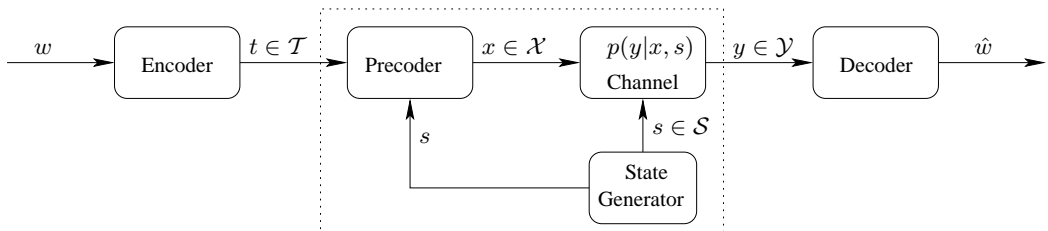}
\caption{The \emph{associated} regular DMC.} \label{J2-fig2}
\end{figure}

The SD-DMC can be used in the way shown in fig. \ref{J2-fig2} to
transmit information. A precoder is added in front of the SD-DMC. A
message $w$ is mapped into $\mathcal{T}^n$, where $\mathcal{T}$ is a
new alphabet. The output of the precoder ranges over $\mathcal{X}$
and depends on the current interference symbol. The regular (without
state) channel from $T$ to $Y$ is defined by the transition
probabilities
\begin{equation} \label{nostate}
q(y|t) = \sum_{s \in \mathcal{S}} p(s) p(y|x=t(s),s),
\end{equation}
where $p(s)$ is the probability of the state $s$. The DMC defined in
(\ref{nostate}) is called the \emph{associated} channel. The codes
for the \emph{associated} channel describe the codes for the SD-DMC
that use only the current state symbols in the encoding operation.
In order to describe all coding schemes for the SD-DMC that use only
the current state symbol in the encoding process, $\mathcal{T}$ must
include all functions from the state alphabet to the input alphabet
of the state-dependent channel. There are a total of $
|\mathcal{X}|^{|\mathcal{S}|} $ of such functions, where $|.|$
denotes the cardinality of a set. Any of the functions can be
represented by a $ |\mathcal{S}| $-tuple $
(x_{1},x_{2},\ldots,x_{{|\mathcal{S}|}}) $ composed of elements of
$\mathcal{X}$, implying that the value of the function at state $ s
$ is $ x_{s}, s=1,2,\ldots,|\mathcal{S}| $.

\section{The Channel Model} \label{chanmodel}
We consider data transmission over the channel
\begin{equation} \label{chmodel}
Y = X + S + N,
\end{equation}
where $ X $ is the channel input, which takes on values in a real
finite set $\mathcal{X}$, $Y$ is the channel output, $N$ is additive
white Gaussian noise with power $\sigma^2$, and the interference $S$
is a discrete random variable that takes on values in a real finite
set $\mathcal{S}$. The sequence of i.i.d. interference symbols is
known causally at the encoder.

The above channel can be considered as a special case of the
state-dependent channel considered by Shannon with one exception,
that the channel output alphabet is continuous. In our case, the
likelihood function $f_{Y|X,S}(y|x,s)$ is used instead of the
transition probabilities. We denote the input to the
\emph{associated} channel by $T$, which can be considered as a
function from $\mathcal{S}$ to $\mathcal{X}$. We denote the
cardinality of $\mathcal{X}$ and $\mathcal{S}$ by $M$ and $Q$,
respectively. Then the cardinality of $\mathcal{T}$ will be $M^Q$,
which is the number all functions from $\mathcal{S}$ to
$\mathcal{X}$.

The likelihood function for the \emph{associated} channel is given
by
\begin{eqnarray}  \label{assoc} \nonumber
f_{Y|T}(y|t) & = & \sum_{s \in \mathcal{S}} p(s)
f_{Y|X,S}(y|t(s),s)                                       \\
& = & \sum_{s \in \mathcal{S}} p(s) f_{N}(y-t(s)-s),
\end{eqnarray}
where $p(s)$ is the probability of the interference symbol $s$ and
$f_N$ denotes the pdf of the Gaussian noise $N$.

Although in this work, we consider a fixed channel input alphabet
$\mathcal{X}$, the transmitted power is not fixed in general. In
fact, for probability distribution $p(s)$ on $\mathcal{S}$ and for a
given coding scheme for the \emph{associated} channel which induces
probability distribution $p(t)$ on the symbols of $\mathcal{T}$, the
transmitted power is given by
\begin{eqnarray} \nonumber
E[X^2] & = & \sum_{t \in {\mathcal{T}}} \sum_{s \in {\mathcal{S}}}
p(t) p(s) E[X^2|t,s] \\ & = & \sum_{t \in {\mathcal{T}}} \sum_{s \in
{\mathcal{S}}} p(t) p(s) t^2(s).
\end{eqnarray}
Thus, in general, the transmitted power depends on the probability
distribution on the interference alphabet. The binary-input channel
with $\mathcal{X}=\{-x,x\}$ is an exception, however, for which we
have $t^2(s)=x^2$ for all $s \in \mathcal{S}$. Therefore, for any
coding scheme and any probability distribution on the interference
alphabet, the transmitted power is equal to $x^2$.

In this work, we do not impose any constraint on the power of the
transmitted signal. However, in the performance comparisons given in
sections \ref{binary} and \ref{Mary} for different scenarios, we
ensure that the transmitted power is the same in all scenarios.

\section{The Code Design Criterion} \label{code}
Any coding scheme for the \emph{associated} channel defined by
(\ref{assoc}) translates to a coding scheme for the actual channel
defined by $f_{Y|X,S}(y|x,s)$. We use the pairwise error probability
(PEP) approach to derive the code design criterion at high SNR.
Since in this work, we consider fixed channel input and interference
alphabets, the high SNR scenario is realized by making the noise
power $\sigma^2$ sufficiently small. This is equivalent to scale up
the transmitted signal and the interference by the same factor for a
given noise power.

Suppose that the messages $w_1$ and $w_2$ are encoded into codewords
$t_1^n \equiv t_1 t_2 \ldots t_n$ and $r_1^n \equiv r_1 r_2 \ldots
r_n$, respectively, where $t_i$ and $r_i$ belong to the alphabet
$\mathcal{T}$, $i=1,\ldots,n$. In the absence of noise, transmission
of the codeword $t_1^n$ can result in many different received
sequences at the channel output depending on the interference
sequence $s_1^n \equiv s_1 s_2 \ldots s_n$. In specific, all
sequences in $\{(t_1(s_1)+s_1, t_2(s_2)+s_2,\ldots, t_n(s_n)+s_n):
s_1^n \in \mathcal{S}^n\}$ represent the transmitted codeword
$t_1^n$ at the channel output. On the other hand, all sequences in
$\{(r_1(s_1)+s_1, r_2(s_2)+s_2,\ldots, r_n(s_n)+s_n): s_1^n \in
\mathcal{S}^n\}$ represent the codeword $r_1^n$. Using maximum
likelihood decoding, the probability of the event that message $w_2$
is decoded given message $w_1$ was sent is given by
\begin{eqnarray} \label{PEP1} \nonumber
\mbox{Pr}\{w_1 \rightarrow w_2|w_1\} & = & \sum_{s_1^n} p(s_1^n)
\mbox{Pr}\{w_1 \rightarrow w_2|w_1,s_1^n\}             \\ \nonumber
& = & \sum_{s_1^n} p(s_1^n) \mbox{Pr}\left\{f_{Y|T}(y_1^n|t_1^n)
\leq f_{Y|T}(y_1^n|r_1^n) |w_1,s_1^n\right\}
\\ \nonumber & = & \sum_{s_1^n}
p(s_1^n) \mbox{Pr}\left\{\prod_{i=1}^{n} f_{Y|T}(y_i|t_i) \leq
\prod_{i=1}^{n} f_{Y|T}(y_i|r_i) |w_1,s_1^n\right\} \\ \nonumber & =
& \sum_{s_1^n} p(s_1^n) \mbox{Pr}\left\{\prod_{i=1}^{n}
\sum_{s \in \mathcal{S}} p(s) f_N(y_i-t_i(s)-s) \leq  \right. \\
& & \hspace{67pt} \left. \prod_{i=1}^{n} \sum_{s \in \mathcal{S}}
p(s) f_N(y_i-r_i(s)-s) |w_1,s_1^n\right\}.
\end{eqnarray}
In appendix \ref{app1}, we have shown that the above error
probability at high SNR is given by
\begin{equation} \label{criterion}
\mbox{Pr}\{w_1 \rightarrow w_2|w_1\} =
O\left(Q\left(\frac{\sqrt{\sum_{i=1}^{n}
d_{\textsf{SI}}^2(t_i,r_i)}}{2\sigma} \right)\right),
\end{equation}
where
\begin{equation}
Q(x) = \int_{x}^{\infty} \frac{1}{\sqrt{2\pi}}
\exp\left(-\frac{y^2}{2}\right) dy,
\end{equation}
and $d_{\textsf{SI}}(t,r)$ (\textsf{SI} stands for side
information), the distance between two input symbols of the
\emph{associated} channel $t$ and $r$, is defined as
\begin{equation} \label{distance}
d_{\textsf{SI}}(t,r) = \min_{s_1,s_2 \in \mathcal{S}}
|t(s_1)+s_1-r(s_2)-s_2|.
\end{equation}
According to (\ref{criterion}), at high SNR, the code design
criterion is to maximize the minimum distance between the codewords
with the distance measure defined in (\ref{distance}).

\subsection{No Side Information at the Encoder - A Comparison}
In order to see how the knowledge of interference at the encoder can
result in larger distances between codewords, consider the channel
model introduced in section \ref{chanmodel} with the exception that
the interference sequence is not known at the encoder. In this case,
the discrete interference is considered as noise. In order to obtain
the PEP for this channel, suppose that messages $v_1$ and $v_2$ are
encoded into $x_1^n \equiv x_1 \cdots x_n \in \mathcal{X}^n$ and
$z_1^n \equiv z_1 \cdots z_n \in \mathcal{X}^n$, respectively.
Similarly, it can be shown that the PEP at high SNR is given by
\begin{equation} \label{critnosi}
\mbox{Pr}\{v_1 \rightarrow v_2|v_1\} =
O\left(Q\left(\frac{\sqrt{\sum_{i=1}^{n} d^2(x_i,z_i)}}{2\sigma}
\right)\right),
\end{equation}
where $d(x,z)$, the distance between two symbols $x$ and $z$ of
$\mathcal{X}$ is defined as
\begin{equation} \label{dis}
d(x,z) = \min_{s_1,s_2 \in \mathcal{S}} |x+s_1-z-s_2|.
\end{equation}
Comparing (\ref{distance}) and (\ref{dis}), it becomes clear that
larger distances among codewords are possible for the channel with
side information at the encoder. In fact, the distance $d(x,z)$ is
equal to $d_{\textsf{SI}}(t,r)$ for $t=(x,\ldots,x)$ and
$r=(z,\ldots,z)$. However, $\mathcal{T}$ has many other symbols,
which may yield larger distances. For example, consider the channel
with $\mathcal{X}=\mathcal{S}=\{-1,+1\}$. For the case without side
information at the encoder, we can compute the distances between
symbols of $\mathcal{X}$ according to (\ref{dis}) as
$d(1,1)=d(-1,-1)=d(1,-1)=0$. Hence, according to (\ref{critnosi}),
it is impossible to transmit data over this channel with low error
probability even at high SNR. For the case with side information at
the encoder, the four symbols of the \emph{associated} channel can
be represented as $u_1=(-1,+1), u_2=(+1,-1), u_3=(+1,+1),
u_4=(-1,-1)$. Using (\ref{distance}), it is easy to check that the
distances between all pairs of the symbols are zero except for
$d_{\textsf{SI}}(u_1,u_2)$ which is $2$. As will be seen in section
\ref{binary}, $u_1$ and $u_2$ can be used in the encoding to achieve
arbitrarily low error probabilities as SNR increases.

It is worth mentioning that the distance measures defined in
(\ref{distance}) or (\ref{dis}) do not satisfy the triangle
inequality. For example, again consider the channel with
$\mathcal{X}=\mathcal{S}=\{-1,+1\}$. The distances between all pairs
of the input symbols of the \emph{associated} channel are zero
except for $d_{\textsf{SI}}(u_1,u_2)$ which is $2$. Therefore, the
triangle inequality does not hold for $d_{\textsf{SI}}(u_1,u_3)$,
$d_{\textsf{SI}}(u_3,u_2)$, and $d_{\textsf{SI}}(u_1,u_2)$.

\section{The Binary Channel} \label{binary}
We call the channel introduced in (\ref{chmodel}) a \emph{binary
channel} when the channel accepts binary input, i.e., $M=2$. There
is no constraints on the cardinality of the interference alphabet.
For the binary channel, the size of $\mathcal{T}$ is $2^Q$. However,
we may not need to use all the symbols of the alphabet in the
encoding. In this section, we show that it is sufficient to use only
two symbols of $\mathcal{T}$ in the encoding as far as the distance
spectrum of the code is concerned. We begin with the following lemma
for the binary channel.

\begin{lemma} \label{lem1}
For the binary channel, there exist at least two symbols in
$\mathcal{T}$ with nonzero distance.
\end{lemma}
\begin{proof}
We may explicitly denote the channel input and interference
alphabets by $\mathcal{X}=\{x_1, x_2\}$ and $\mathcal{S}=\{s_1,
\ldots, s_Q\}$, where $x_1<x_2$ and $s_1<s_2<\cdots<s_Q$. From the
definition of distance in (\ref{distance}), it is sufficient to show
that there exist two elements $t$ and $r$ in $\mathcal{T}$ such that
the corresponding multi-sets \footnote{A multi-set differs from a
set in that each member may have a multiplicity greater than one.
For example, $\{1,3,3,7\}$ is a multi-set of size four where $3$ has
multiplicity two.} (of size $Q$) $\{t(s_1)+s_1,\ldots,t(s_Q)+s_Q\}$
and $\{r(s_1)+s_1,\ldots,r(s_Q)+s_Q\}$ are disjoint. We prove this
by induction on $Q$.

The statement of the lemma holds for $Q=1$ since we may take
$t=(x_1)$ and $r=(x_2)$. Then the sets $\{x_1+s_1\}$ and
$\{x_2+s_1\}$ are disjoint. Now suppose that the statement of the
lemma is true for some $Q$. Therefore, the exist two $Q$-tuples
composed of elements of $\mathcal{X}$ (two input symbols of the
\emph{associated} channel) such that the corresponding multi-sets
are disjoint. We prove that the statement of the lemma hold for
$Q+1$.

The element $x_2+s_{Q+1}$ is larger than any element of the two
multi-sets (of size $Q$). Hence, it does not belong to any of the
multi-sets. If $x_1+s_{Q+1}$ does not belong to any of the
multi-sets too, then we can include the new elements $x_1+s_{Q+1}$
and $x_2+s_{Q+1}$ in the multi-sets of size $Q$ arbitrarily (one
elements in each multi-set). The resulting multi-sets of size $Q+1$
will be disjoint. If $x_1+s_{Q+1}$ belongs to one of the multi-set
of size $Q$, we include it in that multi-set and include
$x_2+s_{Q+1}$ in the other multi-set to form the new disjoint
multi-sets of size $Q+1$. The two $(Q+1)$-tuples (the two input
symbols of the \emph{associated} channel) are then obtained from the
two multi-sets of size $Q+1$ by subtracting the interference symbols
from their elements.
\end{proof}

Lemma \ref{lem1} is in fact a special case of theorem 2 in
\cite{J1}, which was stated in the context of capacity.

Let $u_1$ and $u_2$ be two input symbols of the \emph{associated}
channel with the maximum distance among all pairs of input symbols
of the \emph{associated} channel. Since $d_{\textsf{SI}}(u_1,u_2)>0$
(according to Lemma \ref{lem1}), we have $u_1(s) \neq u_2(s),
\forall s \in \mathcal{S}$, otherwise, from (\ref{distance}),
$d_{\textsf{SI}}(u_1,u_2)=0$. We choose an arbitrary interference
symbol $s \in \mathcal{S}$ to partition $\mathcal{T}$ as follows. We
put $t \in \mathcal{T}$ in $\mathcal{T}_1$ if $t(s)=u_1(s)$,
otherwise (i.e., $t(s)=u_2(s)$) we put $t$ in $\mathcal{T}_2$. Note
that the distance between any two symbols in $\mathcal{T}_j$ is
zero, $j=1,2$.

Suppose that a codebook is designed for the binary channel with
codewords composed of elements of $\mathcal{T}$. We construct a new
codebook from the original one by replacing the elements of the
codewords that belong to $\mathcal{T}_1$ by $u_1$ and replacing the
elements of the codewords that belong to $\mathcal{T}_2$ by $u_2$.
Since the codewords of the new codebook are composed of just two
elements, we may call the new code a binary code.
\begin{theorem} \label{theo1}
The distance spectrum of the binary code constructed by the
procedure described above is at least as good as the distance
spectrum of the original code.
\end{theorem}
\begin{proof}
Consider any two codewords $(t_1,\ldots,t_n)$ and $(r_1,\ldots,r_n)$
from the original codebook, where $t_i,r_i \in \mathcal{T}$. The
squared distance between the two codewords is equal to
$\sum_{i=1}^{n} d_{\textsf{SI}}^2(t_i,r_i)$. For any $i \in
\{1,2,\ldots,n \}$, we consider two cases:

\emph{Case 1}: $t_i$ and $r_i$ belong to the same partition. Then
$d_{\textsf{SI}}(t_i,r_i)=0$, so the replacement will not change the
distance.

\emph{Case 2}: $t_i$ and $r_i$ belong to different partitions. Then
since $d_{\textsf{SI}}(t_i,r_i) \leq d_{\textsf{SI}}(u_1,u_2)$, the
replacement will not decrease the distance.
\end{proof}

According to theorem \ref{theo1}, as far as the distance spectrum of
the code in concerned, it is sufficient to use two symbols of
$\mathcal{T}$ with the maximum distance, namely $u_1$ and $u_2$, in
the encoding for a binary channel. Since $\mathcal{T}$ has size
$2^Q$ for the binary channel, a brute-force search for finding two
symbols in $\mathcal{T}$ with the maximum distance will have
exponential complexity with respect to $Q$. We have proposed an
algorithm with polynomial complexity for finding two symbols with
the maximum distance in appendix \ref{app2}.

Since it is sufficient to use $u_1$ and $u_2$ in the encoding for
the binary channel, we can define the Hamming distance between any
two codewords, which is the number of positions at which the two
codewords are different. Consider two codewords
$c_1=(t_1,\ldots,t_n)$ and $c_2=(r_1,\ldots,r_n)$ with elements from
the binary set $\{u_1,u_2\}$. The squared distance between these
codewords is given by
\begin{equation} \label{d_SI}
\sum_{i=1}^{n} d_{\textsf{SI}}^2(t_i,r_i) =
d_{\textsf{SI}}^2(u_1,u_2) d_H(c_1,c_2),
\end{equation}
where $d_H(c_1,c_2)$ is the Hamming distance between $c_1$ and
$c_2$. Therefore, the problem of designing codes for the binary
channel where the interference sequence is known causally at the
encoder reduces to the design of codes for the binary symmetric
channel. The only difference is that the coding is over the set
$\{u_1,u_2\}$ rather than $\{0,1\}$.

\subsection{Comparison with the Interference-Free Channel}
If we were to use a binary code for the interference-free binary
channel with the input alphabet $\mathcal{X}=\{x_1,x_2\}$, then the
Euclidean distance between any two codewords $c_1$ and $c_2$ of
length $n$ for the interference-free channel would be
\begin{equation} \label{d_E}
d_{E}^2(c_1,c_2) = (x_1-x_2)^2 d_H(c_1,c_2),
\end{equation}
where $d_E$ denotes the Euclidean distance.

Using (\ref{d_SI}) and (\ref{d_E}), we can compare the performance
of a zero-one binary code for the binary channel with causal side
information at the encoder with the same zero-one binary code for
the interference-free binary channel. In the case of channel with
side information, zero and one are mapped to $u_1$ and $u_2$, and in
the case of the interference-free channel, zero and one are mapped
to $x_1$ and $x_2$, respectively. Note that $u_1$ and $u_2$ are
functions from the interference alphabet $\mathcal{S}$ to the
channel input alphabet $\mathcal{X}=\{x_1,x_2\}$.

It is clear from (\ref{distance}) that
\begin{equation} \label{ineq}
d_{\textsf{SI}}(u_1,u_2) \leq |x_1-x_2|.
\end{equation}
Therefore, using (\ref{d_SI}) and (\ref{d_E}), the distance spectrum
of the code for the interference-free channel is at least as good as
the distance-spectrum of the code for the channel with known
interference at the encoder. Of course, this is not surprising.
However, it is interesting to search for the conditions that
(\ref{ineq}) is satisfied with equality.

If (\ref{ineq}) is satisfied with equality, the distance spectrum of
the two codes will be the same. In other words, if (\ref{ineq}) is
satisfied with equality, the knowledge of interference at the
encoder enables us to achieve the same performance (in terms of
order of probability of error) as the interference-free case at high
SNR.

We may explicitly denote the interference alphabet by
$\mathcal{S}=\{s_1, \ldots, s_Q\}$, where $s_1<s_2<\cdots<s_Q$. Then
the following theorem holds.
\begin{theorem} \label{theo2}
$d_{\textsf{SI}}(u_1,u_2) = |x_1-x_2|$ if and only if
\begin{displaymath}
\min_{i \neq j } |s_i-s_j| \geq |x_1-x_2|.
\end{displaymath}
\end{theorem}
\begin{proof}
If $\min |s_i-s_j| \geq |x_1-x_2|$, we may take
$u_1=(x_1,x_2,x_1,\ldots)$ and $u_2=(x_2,x_1,x_2,\ldots)$. Then we
have
\begin{eqnarray} \nonumber
d_{\textsf{SI}}(u_1,u_2) & = & \min_{i,j}
|u_1(s_i)+s_i-u_2(s_j)-s_j| \\ \nonumber & = & \min
\left\{|x_1+s_k-x_2-s_k|, |x_1+s_{2k_1+1}-x_2-s_{2k_2+1}|_{k_1 \neq
k_2} \right.\\ \nonumber & & \qquad \left.
|x_1+s_{2k_1+1}-x_1-s_{2k_2}|_{k_1,k_2},
|x_2+s_{2k_1}-x_2-s_{2k_2+1}|_{k_1,k_2} \right\} \\ \nonumber & = &
\min \left\{|x_1-x_2|, |x_1+s_{2k_1+1}-x_2-s_{2k_2+1}|_{k_1 \neq
k_2}, |s_{2k_1+1}-s_{2k_2}|_{k_1,k_2} \right\}.\\
\end{eqnarray}
We also have
\begin{eqnarray} \nonumber
|x_1+s_{2k_1+1}-x_2-s_{2k_2+1}| & \geq &
|s_{2k_1+1}-s_{2k_2+1}|-|x_1-x_2| \\ \nonumber & \geq & 2 \min
|s_i-s_j| - |x_1-x_2| \qquad \mbox{for} \, \; k_1 \neq k_2 \\
& \geq & |x_1-x_2|
\end{eqnarray}
and
\begin{eqnarray} \nonumber
|s_{2k_1+1}-s_{2k_2}| & \geq & \min |s_i-s_j| \quad \forall\; k_1,
k_2
\\ & \geq & |x_1-x_2|.
\end{eqnarray}
Therefore, $d_{\textsf{SI}}(u_1,u_2) = |x_1-x_2|$.

For the other direction, suppose that $\min |s_i-s_j| < |x_1-x_2|$.
We will show that $d_{\textsf{SI}}(u_1,u_2) < |x_1-x_2|$. Suppose
that $s_k, s_{k+1} \in \mathcal{S}$ achieve the minimum of
$|s_i-s_j|$ and $t_1$ and $t_2$ are arbitrary elements of
$\mathcal{T}$. We consider two non-trivial cases:

\emph{Case 1}: $t_1(s_k)=t_1(s_{k+1})=x_1$ and
$t_2(s_k)=t_2(s_{k+1})=x_2$. Then $d_{\textsf{SI}}(t_1,t_2) \leq
|t_1(s_{k+1})+s_{k+1}-t_2(s_k)-s_k| < |x_1-x_2|$.

\emph{Case 2}: $t_1(s_k)=x_1, t_1(s_{k+1})=x_2$ and $t_2(s_k)=x_2,
t_2(s_{k+1})=x_1$. Then $d_{\textsf{SI}}(t_1,t_2) \leq
|t_1(s_k)+s_k-t_2(s_{k+1})-s_{k+1}| < |x_1-x_2|$.
\end{proof}

\begin{figure}[t]
\centering
\includegraphics[scale = 0.8]{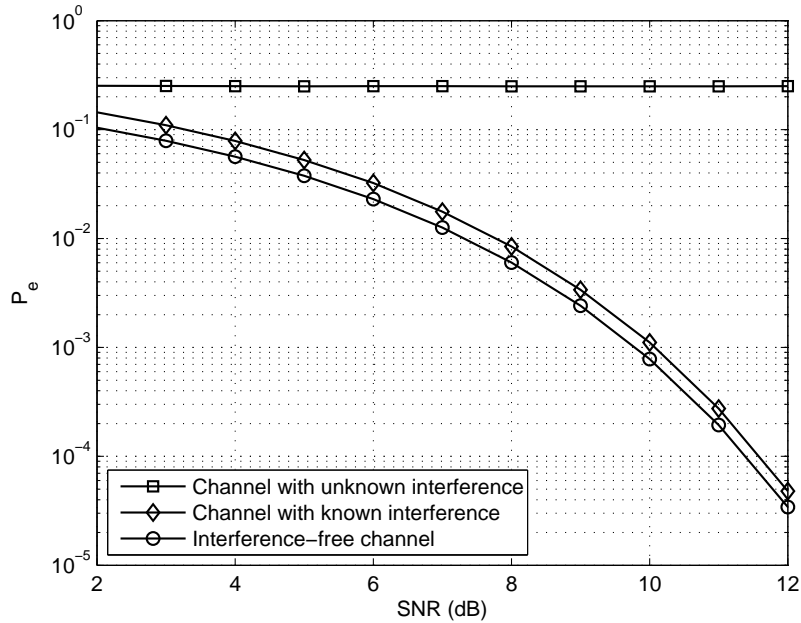}
\caption{Error probability vs. SNR for the binary input AWGN channel
with/without known/unknown interference.
$\mathcal{X}=\mathcal{S}=\{-1,+1\}$.} \label{J2-fig3}
\end{figure}

As an example, consider a binary channel with
$\mathcal{X}=\mathcal{S}=\{-1,+1\}$ and equiprobable interference
symbols. The two symbols with the maximum distance in the input
alphabet of the \emph{associated} channel are
$u_1=(-1,+1),u_2=(+1,-1)$. We have simulated the error probability
performance of the above uncoded system with maximum likelihood
decoding. The error probability vs. SNR
$\left(=\frac{1}{\sigma^2}\right)$ for the above channel is plotted
in fig. \ref{J2-fig3}. The error probability curve for the
interference-free channel with $\mathcal{X}=\{-1,+1\}$ is plotted
for comparison. For the interference-free channel,
$P_e=Q(\frac{1}{\sigma})$. It is easy to check that in this example,
$d_{\textsf{SI}}(u_1,u_2)=|x_1-x_2|=2$. As it can be seen, the error
probability curves decay at the same rate with increasing SNR as
expected. The error probability curve for the scenario that the
interference is not known at the encoder, is plotted for comparison.
In this scenario, the error probability curve reaches an error floor
of $\frac{1}{4}$.

\begin{figure}[t]
\centering
\includegraphics[scale = 0.8]{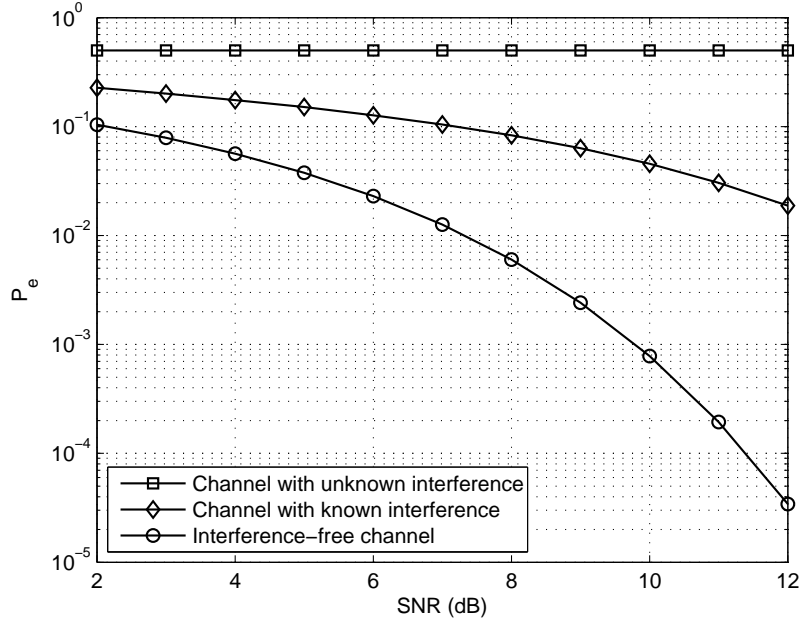}
\caption{Error probability vs. SNR for the binary input AWGN channel
with/without known/unknown interference. $\mathcal{X}=\{-1,+1\}$,
$\mathcal{S}=\{-1,0,+1\}$.} \label{J2-fig4}
\end{figure}

Another example is illustrated in fig. \ref{J2-fig4}. For this
example, $\mathcal{X}=\{-1,+1\}, \mathcal{S}=\{-1,0,+1\}$. We can
find by inspection two symbols of the \emph{associated} channel
input alphabet with the maximum distance as
$u_1=(-1,-1,+1),u_2=(+1,+1,-1)$. Here, we have
$d_{\textsf{SI}}(u_1,u_2)=1 < |x_1-x_2|=2$. Therefore, the error
probability curve for the channel with known interference at the
encoder does not decay as fast as the error probability curve for
the interference-free channel. For the scenario that the
interference is not known at the encoder, the error probability
curve reaches an error floor of $\frac{1}{6}$.

\section{The $M$-ary Channel} \label{Mary}
In general, the statement of theorem \ref{theo1} is not extendable
to the case with $M>2$ channel input symbols. In fact, by using more
than $M$ input symbols of the \emph{associated} channel, we can
obtain a better codebook in terms of distance spectrum than any
other codebook composed of just $M$ input symbols of the
\emph{associated} channel. An example showing this is given in
appendix \ref{app3}. However, under some condition on the channel
input and interference alphabets, the statement of theorem
\ref{theo1} can  be generalized to the case with $M>2$.

\begin{theorem} \label{theo3}
As far as the distance spectrum of code is concerned, it is
sufficient to use $M$ (out of $M^Q$) input symbols of the
\emph{associated} channel in the encoding if
\begin{displaymath}
\min_{s_i, s_j \in \mathcal{S}} |s_i-s_j| \geq 2 \max_{x_i,x_j \in
\mathcal{X}} |x_i-x_j|.
\end{displaymath}
\end{theorem}
\begin{proof}
Consider the $M$ input symbols of the \emph{associated} channel
$u_1=(x_1,\ldots,x_1)$, $u_2=(x_2,\ldots,x_2)$, $\ldots$,
$u_M=(x_M,\ldots,x_M)$. We use these symbols to partition the
\emph{associated} channel input alphabet $\mathcal{T}$ as follows.
Put $t \in \mathcal{T}$ in $\mathcal{T}_i$ if the first element of
$t$ is $x_i$, $i=1,2,\ldots,M$. Note that $\mathcal{T}_i$ has size
$M^{Q-1}$ and the distance between any two symbols in
$\mathcal{T}_i$ is zero, $i=1,2,\ldots,M$. For any $p, q=1,\ldots,
M$, we have
\begin{eqnarray} \nonumber
d_{\textsf{SI}}(u_p,u_q) & = & \min_{k_1,k_2}
|x_p+s_{k_1}-x_q-s_{k_2}| \\ & = & \min
\left\{|x_p-x_q|,|x_p+s_{k_1}-x_q-s_{k_2}|_{k_1 \neq k_2} \right\}.
\end{eqnarray}
We also have
\begin{eqnarray} \nonumber
|x_p+s_{k_1}-x_q-s_{k_2}| & \geq & |s_{k_1}-s_{k_2}|-|x_p-x_q|
\\ \nonumber & \geq & 2 \max |x_i-x_j| - |x_p-x_q| \quad \mbox{for}
\; k_1 \neq k_2 \\ & \geq & |x_p-x_q|,
\end{eqnarray}
Therefore, $d_{\textsf{SI}}(u_p,u_q) = |x_p-x_q|$. Note that the
distance between any two symbols from $\mathcal{T}_p$ and
$\mathcal{T}_q$ is at most $|x_p-x_q|=d_{\textsf{SI}}(u_p,u_q)$.

Suppose that a codebook is designed with codewords composed of
possibly all elements of $\mathcal{T}$. We construct a new codebook
from the original one by replacing the elements of the codewords
that belong to $\mathcal{T}_i$ by $u_i$, $i=1,2,\ldots,M$. It is
easy to check that the distance spectrum of the new code is at least
as good as the distance spectrum of the original code.
\end{proof}

According to theorem \ref{theo3}, it is sufficient to use only the
symbols $u_1,\ldots,u_M$ in the encoding. But any of these symbols
is a constant function from $\mathcal{S}$ to $\mathcal{X}$.
Therefore, the same symbol enters the channel regardless of the
current interference symbol. This suggests that the knowledge of
interference symbols at the encoder is not helpful in terms of
distance spectrum improvement provided that the condition of theorem
\ref{theo3} is satisfied. In fact, with the condition of theorem
\ref{theo3}, we have
\begin{equation}
d_{\textsf{SI}}(u_i,u_j) = d(x_i,x_j) = d_E(x_i,x_j), \qquad i,j =
1, \ldots, M.
\end{equation}
where $d(.,.)$, defined in (\ref{dis}), is the distance measure when
the interference is not known at the encoder and $d_E(.,.)$ is the
Euclidean distance measure. Therefore, the error probability
performance of a code for the channel with known/unknown
interference at the encoder will be the same as the performance of
the same code for the interference-free channel at high SNR.

It is worth mentioning that for the above-mentioned three scenarios
the codes for the interference-free channel, the channel with known
interference at the encoder, and the channel with unknown
interference use the same transmitted power.

\section{A More General Channel Model} \label{general}
Although we have considered the AWGN channel with additive
interference so far, our treatment applies to more general channels
characterized by
\begin{equation} \label{general_chan}
Y = f(X,S)+ N,
\end{equation}
where $f$ is an arbitrary function of two variables, $S$ is the
channel state which is known causally at the encoder, $X$ is the
channel input, and $N$ is white Gaussian noise. Another special case
of this more general channel is the fast fading channel
\begin{equation}
Y = SX + N,
\end{equation}
where $S$ is the fading coefficient. For the general channel model
(\ref{general_chan}), the distance between two symbols $t$ and $r$
of $\mathcal{T}$ is defined as
\begin{equation}
d_{\textsf{SI}}(t,r) = \min_{s_1,s_2\in \mathcal{S}}
|f(t(s_1),s_1)-f(t(s_2),s_2)|.
\end{equation}

Theorem \ref{theo1} on the binary channel also holds for the general
channel model. However, the maximum distance among pairs of symbols
of $\mathcal{T}$ may be zero; i.e., lemma \ref{lem1} does not hold
true in general. Theorems \ref{theo2} and \ref{theo3} do not hold
for the more general channel model in (\ref{general_chan}) and are
specific to the AWGN with additive interference channel model.

\section{Conclusion} \label{conclude}
In this paper, we derived the code design criterion at high SNR for
the $M$-ary input AWGN channel with additive $Q$-level interference,
where the sequence of interference symbols is known causally at the
encoder. The code design is over an input alphabet $\mathcal{T}$ of
size $M^Q$. The performance of a code for our channel at high SNR is
governed by the minimum distance between the codewords with elements
from $\mathcal{T}$. We may not need to use all symbols of
$\mathcal{T}$ in the encoding. In particular, we showed that for the
case $M=2$, as far as the distance spectrum of the code is
concerned, we just need to use two symbols of $\mathcal{T}$ with the
maximum distance among all pairs of symbols. This reduces the code
design problem for our channel to code design for binary symmetric
channel which has been well researched in the literature.

\appendices
\section{Derivation of Code Design Criterion at high
SNR}\label{app1} Define
\begin{eqnarray}
\mathcal{A}_i & = & \{t_i(s)+s:s\in\mathcal{S}\}, \hspace{50pt}
i=1,\ldots,n,\\ \mathcal{B}_i & = & \{r_i(s)+s:s\in\mathcal{S}\},
\hspace{50pt} i=1,\ldots,n.
\end{eqnarray}
It is worth mentioning that the cardinality of $\mathcal{A}_i$ (or
$\mathcal{B}_i$) can be less than $Q$, $i=1,\ldots,n,$ since
different interference symbols may yield the same element in
$\mathcal{A}_i$ (or $\mathcal{B}_i$). For any $i=1,\ldots,n$, we
have
\begin{eqnarray}
\sum_{s \in \mathcal{S}} p(s) f_N(y-t_i(s)-s) & = & \sum_{a \in
\mathcal{A}_i} p(a) f_N(y-a),\\ \sum_{s \in \mathcal{S}} p(s)
f_N(y-r_i(s)-s) & = & \sum_{b \in \mathcal{B}_i} p(b) f_N(y-b),
\end{eqnarray}
where $p(a)$ and $p(b)$ are obtained from $p(s)$ according to
\begin{eqnarray}
p(a) & = & \sum_{s \in \mathcal{S} : t_i(s)+s=a} p(s),\\
p(b) & = & \sum_{s \in \mathcal{S} : r_i(s)+s=b} p(s).
\end{eqnarray}

For any sequence $a_1^n \equiv a_1 \cdots a_n \in \mathcal{A}_1
\times \cdots \times \mathcal{A}_n$ and $b_1^n \equiv b_1 \cdots b_n
\in \mathcal{B}_1 \times \cdots \times \mathcal{B}_n$, we define the
events
\begin{eqnarray}
E_1(a_1^n) & = & \bigcap_{i=1}^n \left(a_i = \arg \min_{a \in
\mathcal{A}_i} |y_i-a| \right), \\ E_2(b_1^n) & = & \bigcap_{i=1}^n
\left(b_i = \arg \min_{b \in \mathcal{B}_i} |y_i-b|\right),
\end{eqnarray}
given that $w_1$ has been sent and the interference sequence $s_1^n$
has occurred. The event $E_1(a_1^n)$ simply means that $a_i$ is the
closest point to the received signal $y_i$ (given $w_1$ has been
sent and the interference sequence $s_1^n$ has occurred) among all
points of $\mathcal{A}_i$ for all $i=1,\ldots,n$.

Any term in the error probability in (\ref{PEP1}) can be written as
\begin{eqnarray}  \label{p11} \nonumber
\hspace{-8pt}& & \hspace{-8pt}\mbox{Pr}\left\{\prod_{i=1}^{n}
\sum_{a \in \mathcal{A}_i} p(a) f_N(y_i-a) \leq  \prod_{i=1}^{n}
\sum_{b \in \mathcal{B}_i} p(b) f_N(y_i-b) |w_1,s_1^n\right\} \\
\nonumber \hspace{-8pt}&=& \hspace{-8pt} \sum_{a_1^n} \sum_{b_1^n}
\mbox{Pr}\left\{\prod_{i=1}^{n} \sum_{a \in \mathcal{A}_i} p(a)
f_N(y_i-a) \leq  \prod_{i=1}^{n} \sum_{b \in \mathcal{B}_i} p(b)
f_N(y_i-b), E_1(a_1^n), E_2(b_1^n) |w_1,s_1^n \right\}\\ \nonumber
\hspace{-8pt} & = & \hspace{-8pt} \sum_{a_1^n} \sum_{b_1^n}
\mbox{Pr}\left\{\prod_{i=1}^{n} f_N(y_i-a_i)
\left(p(a_i)+\sum_{\substack{a \in \mathcal{A}_i \\ a \neq a_i}}
p(a) \frac{f_N(y_i-a)}{f_N(y_i-a_i)} \right) \right.
\\ \nonumber \hspace{-8pt} & & \hspace{25pt} \leq \left.
\prod_{i=1}^{n} f_N(y_i-b_i) \left(p(b_i)+\sum_{\substack{b \in
\mathcal{B}_i \\ b \neq b_i}} p(b) \frac{f_N(y_i-b)}{f_N(y_i-b_i)}
\right)
, E_1(a_1^n), E_2(b_1^n) |w_1,s_1^n \right\}\\
\hspace{-8pt}&=&\hspace{-8pt} \sum_{a_1^n} \sum_{b_1^n}
\mbox{Pr}\left\{\sum_{i=1}^n (y_i-a_i)^2 \geq \sum_{i=1}^n
(y_i-b_i)^2 + K \sigma^2, E_1(a_1^n), E_2(b_1^n) |w_1,s_1^n
\right\},
\end{eqnarray}
where $K = K(y_1^n,a_1^n,b_1^n)$ is given by
\begin{equation}
K(y_1^n,a_1^n,b_1^n) = 2 \sum_{i=1}^{n} \log
\frac{p(a_i)+\sum_{\substack{a \in \mathcal{A}_i \\ a \neq a_i}}
p(a) \frac{f_N(y_i-a)}{f_N(y_i-a_i)}}{p(b_i)+\sum_{\substack{b \in
\mathcal{B}_i \\ b \neq b_i}} p(b) \frac{f_N(y_i-b)}{f_N(y_i-b_i)}}.
\end{equation}
Given the events $E_1(a_1^n)$ and $E_2(b_1^n)$, it is easy to check
that $K(y_1^n,a_1^n,b_1^n)$ is bounded as
\begin{equation}
K_1(a_1^n) = 2 \sum_{i=1}^{n} \log p(a_i) < K(y_1^n,a_1^n,b_1^n) <
K_2(b_1^n) = 2 \sum_{i=1}^{n} \log \frac{1}{p(b_i)}.
\end{equation}
As we consider the high SNR regime, we may assume that the noise
power is sufficiently small so that the error probability
(\ref{PEP1}) can be well approximated by
\begin{equation} \label{pe}
\sum_{s_1^n} p(s_1^n) \sum_{a_1^n} \sum_{b_1^n}
\mbox{Pr}\left\{\sum_{i=1}^n (y_i-a_i)^2 \geq \sum_{i=1}^n
(y_i-b_i)^2, E_1(a_1^n), E_2(b_1^n) |w_1,s_1^n \right\}.
\end{equation}
Any term in the summation (\ref{pe}) can be upper bounded as
\begin{eqnarray}  \label{upper} \nonumber
& & \mbox{Pr}\left\{\sum_{i=1}^n (y_i-a_i)^2 \geq \sum_{i=1}^n
(y_i-b_i)^2, E_1(a_1^n), E_2(b_1^n) |w_1,s_1^n \right\} \\ \nonumber
& \leq & \mbox{Pr}\left\{\sum_{i=1}^n (y_i-c_i)^2 \geq \sum_{i=1}^n
(y_i-b_i)^2, E_1(a_1^n), E_2(b_1^n) |w_1,s_1^n \right\} \\ \nonumber
& \leq & \mbox{Pr}\left\{\sum_{i=1}^n (y_i-c_i)^2 \geq \sum_{i=1}^n
(y_i-b_i)^2 |w_1,s_1^n \right\} \\ \nonumber & = & Q\left(
\frac{\sqrt{\sum_{i=1}^{n}|c_i-b_i|^2}}{2 \sigma}\right) \\
& \leq & Q\left(\frac{\sqrt{\sum_{i=1}^{n}
d_{\textsf{SI}}^2(t_i,r_i)}}{2\sigma}\right),
\end{eqnarray}
where
\begin{equation}
c_i = t_i(s_i) + s_i, \quad i=1,\ldots,n.
\end{equation}
The first inequality is due to the fact that given $E_1(a_1^n)$, we
have $|y_i-a_i| \leq |y_i-c_i|, i=1,\ldots,n$.

In the following, we show that the upper bound (\ref{upper}) is
tight for the term(s) in the summation (\ref{pe}) satisfying
\begin{equation} \label{minD}
\{a_i,b_i\} = \arg \min_{\substack{a \in \mathcal{A}_i \\ b \in
\mathcal{B}_i}} |a-b|, \quad i=1,\ldots,n,
\end{equation}
and
\begin{equation} \label{same}
a_i = c_i, \quad i=1,\ldots,n.
\end{equation}

Any term in (\ref{pe}) equals the integral of the joint probability
distribution of $y_1^n \equiv y_1 \cdots y_n$ (given $w_1, s_1^n$)
over the region in the $n$-dimensional Euclidean space defined by
\begin{equation} \label{reg1}
\left\{y_1^n : \sum_{i=1}^n (y_i-a_i)^2 \geq \sum_{i=1}^n
(y_i-b_i)^2, E_1(a_1^n), E_2(b_1^n) \right\}.
\end{equation}

\begin{figure}[t]
\centering
\includegraphics[scale = 0.9]{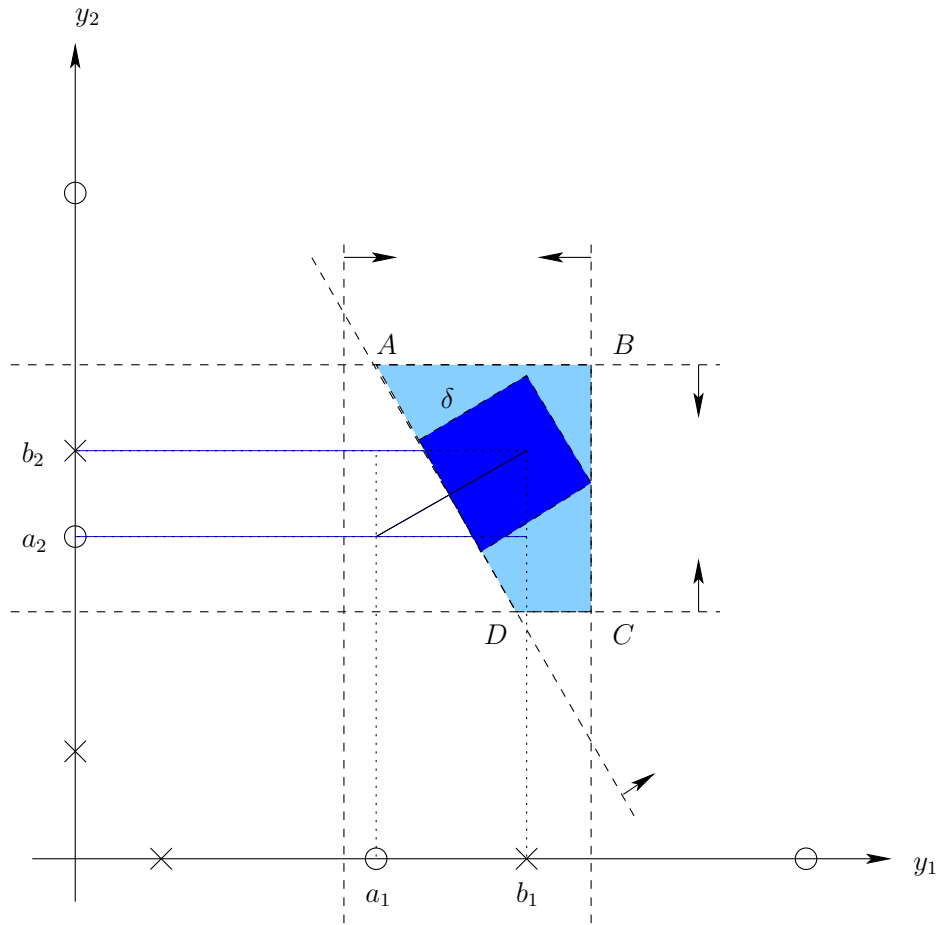}
\caption{Illustrating the regions of integration for dimension
$n=2$.} \label{J2-fig5}
\end{figure}

This region is illustrated by the shaded area ABCD in fig.
\ref{J2-fig5} for $n=2$. The horizontal and vertical boundaries of
ABCD correspond to the events $E_1(a_1^2)$ and $E_2(b_1^2)$. The
elements of $\mathcal{A}_i$ and $\mathcal{B}_i$ are shown by $\circ$
and $\times$, respectively. The other boundary of ABCD which
corresponds to $\sum_{i=1}^2 (y_i-a_i)^2 \geq \sum_{i=1}^2
(y_i-b_i)^2$ is the perpendicular bisector of the line segment
connecting $a_1^2$ to $b_1^2$. We may consider an $n$-cube inside
this region with sides equal to some $\delta > 0$ as shown in fig.
\ref{J2-fig5} and perform the integration over this smaller region
to obtain a lower bound for the term(s) in the summation (\ref{pe})
satisfying (\ref{minD}) and (\ref{same}).

In summary, for the terms in (\ref{pe}) which satisfy (\ref{minD})
and (\ref{same}), we have
\begin{eqnarray}  \label{lower} \nonumber
& & \mbox{Pr}\left\{\sum_{i=1}^n (y_i-a_i)^2 \geq \sum_{i=1}^n
(y_i-b_i)^2, E_1(a_1^n), E_2(b_1^n) |w_1,s_1^n \right\} \\
\nonumber & \geq & \left[
1-Q\left(\frac{\delta}{2\sigma}\right)\right]^{n-1}
\left[Q\left(\frac{\|b_1^n-a_1^n \|}{2\sigma}\right) -
Q\left(\frac{\|b_1^n-a_1^n \|+\delta}{2\sigma}\right)\right]
\\\nonumber & \simeq & Q\left(\frac{\|b_1^n-a_1^n \|}{2\sigma}
\right) \qquad \mbox{as} \, \sigma \rightarrow 0 \\ & = &
Q\left(\frac{\sqrt{\sum_{i=1}^{n}
d_{\textsf{SI}}^2(t_i,r_i)}}{2\sigma} \right),
\end{eqnarray}
where the right hand side of the inequality in (\ref{lower}) equals
the integral of the joint probability distribution of $y_1^n \equiv
y_1 \cdots y_n$ (given $w_1, s_1^n$) over the smaller region, which
is obtained by using the fact that $y_1^n$ is Gaussian centered at
$c_1^n=a_1^n$ and by applying the necessary rotation.

\section{A polynomial complexity algorithm for finding two symbols
of $\mathcal{T}$ with the maximum distance} \label{app2} 
\begin{figure}[t]
\centering
\includegraphics[scale = 0.52]{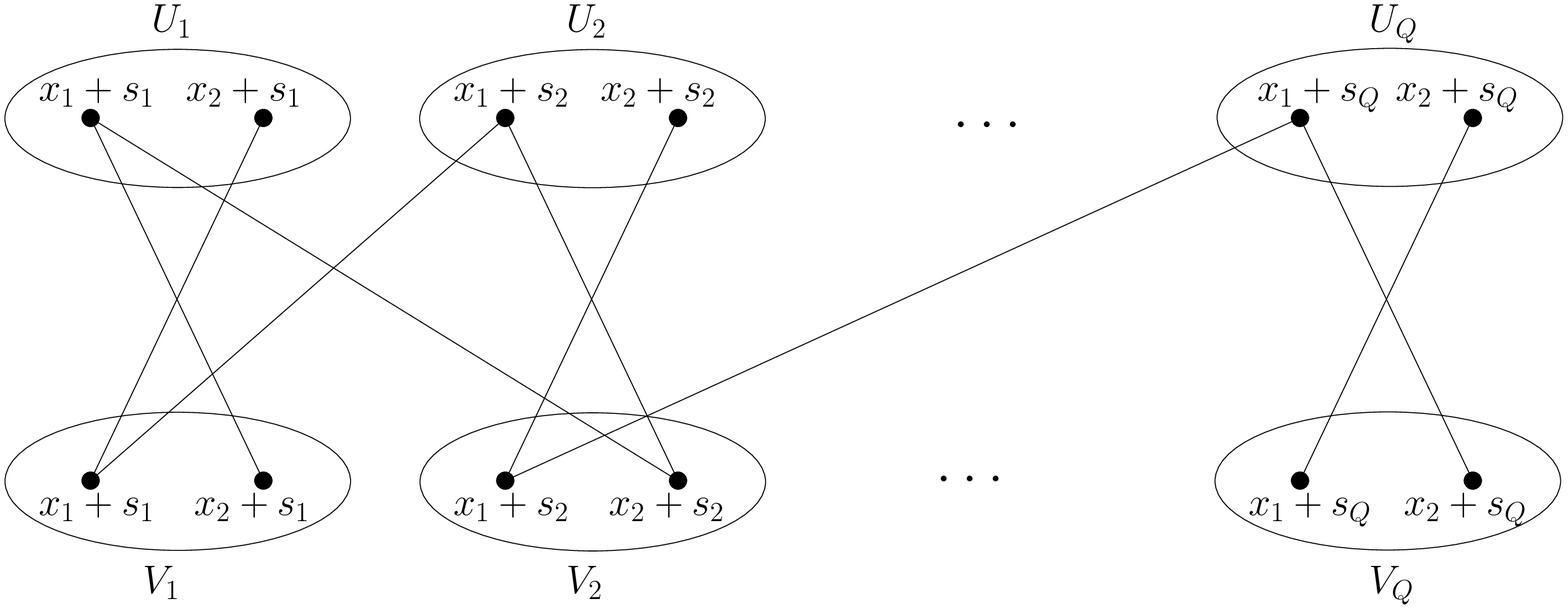}
\caption{Graph representation for the problem of finding two symbols
of $\mathcal{T}$ with the maximum distance.} \label{J2-fig6}
\end{figure}

We propose
an algorithm for finding two symbols of $\mathcal{T}$ with distance
greater than or equal to some $d_0 > 0$. Then we explain how to find
two symbols in $\mathcal{T}$ with the maximum distance. Consider the
bipartite graph $G(U,V,E)$ shown in fig. \ref{J2-fig6} with $2Q$
vertices at each part. Each of the non-intersecting sets
$U_1,\cdots,U_Q$ contains two vertices of the upper part $U$ and
each of the nonintersecting sets $V_1,\cdots,V_Q$ contains two
vertices of the lower part $V$. The vertices of the sets
$U_i=\{u_{i1},u_{i2}\}$ and $V_i=\{v_{i1},v_{i2}\}$ are labeled by
the elements of the set $\mathcal{X}+s_i=\{x_1+s_i,x_2+s_i\}$,
$i=1,\ldots,Q$. A vertex in $U_i$ is connected to a vertex in $V_j$
if the absolute value of the difference of their labels is greater
than or equal to $d_0$, $i,j = 1,\ldots, Q$.

From the definition of distance in (\ref{distance}), there exist two
symbols in $\mathcal{T}$ with distance $d \geq d_0$ if and only if
$G$ has a complete bipartite subgraph $K_{Q,Q}$ with exactly one
vertex in each $U_i$ and each $V_j$. If such a subgraph exists, we
label the edges of the subgraph by $1$ and we label the rest of the
edges of $G$ by $0$. We denote the label of edge $e$ by $y_e \in
\{0,1\}$. Such a labeling satisfies the following set of constraints
\begin{eqnarray} \label{const1}
\sum_{e:e \cap U_i \neq \phi} y_e & = & Q, \qquad i=1,\ldots,Q, \\
\label{const2} \sum_{e:e \cap V_i \neq \phi} y_e & = & Q, \qquad
i=1,\ldots,Q, \\ \label{const3} y_e \in \{0,1\}. & &
\end{eqnarray}
Note that by definition, an edge of a graph is a set of two
vertices. Therefore, the notation $e \cap U_i$ in (\ref{const1}) is
meaningful. The equations (\ref{const1}) and (\ref{const2}) state
that the sum of the labels of the edges going out of any $U_i$ and
$V_i$ is $Q$.

We devise an objective function for the constraints (\ref{const1}),
(\ref{const2}), and (\ref{const3}) such that the objective function
takes a \emph{given} maximum value only for a labeling with label
$1$ for the edges of the subgraph $K_{Q,Q}$ and label $0$ for the
rest of the edges. Consider the following optimization problem
\begin{eqnarray}  \label {opt_int} \nonumber
\max_{y_e} & & \sum_{i=1}^Q \sum_{j=1}^2 \left( \sum_{e: u_{ij} \in
e} y_e \right)^2+\sum_{i=1}^Q \sum_{j=1}^2 \left( \sum_{e: v_{ij}
\in e} y_e \right)^2\\ \nonumber \mbox{subject to} & & \\ \nonumber
& & \sum_{e:e \cap U_i \neq \phi} y_e = Q, \qquad i=1,\ldots,Q, \\
\nonumber & & \sum_{e:e \cap V_i \neq \phi} y_e = Q, \qquad
i=1,\ldots,Q, \\ & & y_e \in \{0,1\}.
\end{eqnarray}
In the following, we find the maximum of the above optimization
problem for the foregoing labeling. Given the constraints of
(\ref{opt_int}), we have
\begin{eqnarray}
\sum_{j=1}^2 \left( \sum_{e: u_{ij} \in e} y_e \right) & = &
\sum_{e:e \cap U_i \neq \phi} y_e = Q, \qquad i=1,\ldots, Q, \\
\sum_{j=1}^2 \left( \sum_{e: v_{ij} \in e} y_e \right) & = &
\sum_{e:e \cap V_i \neq \phi} y_e = Q, \qquad i=1,\ldots, Q.
\end{eqnarray}
If the sum of two nonnegative variables is constant, then the sum of
their squares takes its maximum if one of the variables is zero.
Therefore, for any $i=1,\ldots,Q$, the maximum of
\begin{displaymath}
\sum_{j=1}^2 \left( \sum_{e: u_{ij} \in e} y_e \right)^2
\end{displaymath}
and
\begin{displaymath}
\sum_{j=1}^2 \left( \sum_{e: v_{ij} \in e} y_e \right)^2
\end{displaymath}
will be $Q^2$ and this maximum occurs if and only if one vertex in
any of $U_1, \ldots, U_Q$ and $V_1, \ldots, V_Q$ is connected to $Q$
edges with label $1$ and the other vertex in any of $U_1, \ldots,
U_Q$ and $V_1, \ldots, V_Q$ is not connected to any edge with label
$1$. This is equivalent to the existence of a subgraph $K_{Q,Q}$.
Then the maximum of the objective function in (\ref{opt_int}) will
be $Q \times Q^2 + Q \times Q^2=2Q^3$.

We may relax the integrality constraint (\ref{const3}) and change
equality signs in (\ref{const1}) and (\ref{const2}) to inequality
signs to obtain the following optimization program
\begin{eqnarray}  \label{opt_relax} \nonumber
\max_{y_e} & & \sum_{i=1}^Q \sum_{j=1}^2 \left( \sum_{e: u_{ij} \in
e} y_e \right)^2+\sum_{i=1}^Q \sum_{j=1}^2 \left( \sum_{e: v_{ij}
\in e} y_e \right)^2\\ \nonumber \mbox{subject to} & & \\ \nonumber
& & \sum_{e:e \cap U_i \neq \phi} y_e \leq Q, \qquad i=1,\ldots,Q, \\
\nonumber & & \sum_{e:e \cap V_i \neq \phi} y_e \leq Q, \qquad
i=1,\ldots,Q, \\ & & 0 \leq y_e \leq 1.
\end{eqnarray}
Using the same argument as in the previous paragraph, the value
$2Q^3$ is also achievable for the above maximization problem if and
only if a subgraph $K_{Q,Q}$ of the graph $G$ exists. The above
optimization problem is a \emph{quadratic programming} problem
\cite{Fletcher87} with convex objective function and can be solved
in polynomial time \cite{Khach79} in terms of the number of edges of
$G$, which is at most $4Q^2$.

In summary, we turned the problem of finding two symbols in
$\mathcal{T}$ with distance at least $d_0>0$ into the quadratic
programming problem (\ref{opt_relax}). If the maximum value of
(\ref{opt_relax}) is $2Q^3$, then two such symbols are obtained from
the optimal solution of (\ref{opt_relax}). Otherwise, two such
symbols do not exist.

To find two symbols in $\mathcal{T}$ with the maximum distance, we
need to run the described algorithm for a few values for $d_0$. We
can obtain an upper bound on the number of possible distances
between symbols of $\mathcal{T}$. From the definition of distance in
(\ref{distance}), a loose upper bound is $M^2 Q^2=4Q^2$. By using
the binary search algorithm \cite{Knut97}, the search over possible
distances can be done with logarithmic complexity with respect to
the number of possible distances.

It is worth mentioning that our proposed algorithm can be extended
to find $K \geq 2$ symbols of $\mathcal{T}$ with the maximum minimum
distance among $K$ symbols for the general case $M \geq 2$.

\section{An example that shows using more than $M$ symbols of
$\mathcal{T}$ results in larger minimum distance ($M>2$)}
\label{app3} Consider the channel with $\mathcal{X}=\{1,4,5,7\}$ and
$\mathcal{S}=\{0,4\}$. Consider the following codebook with six
codewords of length two that uses seven symbols of the
\emph{associated} channel.
\begin{eqnarray} \nonumber
\mbox{Codeword 1}: \qquad ((4,1),(5,1))&\\ \nonumber \mbox{Codeword
2}: \qquad ((4,1),(1,5))&\\ \nonumber \mbox{Codeword 3}:
\qquad((5,4),(5,4))&\\ \nonumber \mbox{Codeword 4}:
\qquad((5,4),(4,5))&\\ \nonumber \mbox{Codeword 5}:
\qquad((1,5),(4,1))&\\ \nonumber \mbox{Codeword 6}:
\qquad((1,5),(1,4))
\end{eqnarray}
The minimum distance of the above code is $3$. However, it can be
verified by a computer program that any code for this channel with
codebook size six and length two that uses any four symbols of the
\emph{associated} channel yields a minimum distance less than $3$.

\end{document}